\documentclass[12pt]{elsarticle}

\usepackage{lineno,hyperref}
\usepackage[english]{babel}
\usepackage{amsmath}
\usepackage{amsfonts}
\usepackage{textgreek}
\usepackage{graphicx}
\usepackage{breqn}
\usepackage{undertilde}
\usepackage{appendix}
\usepackage{accents}
\usepackage{enumitem}

\modulolinenumbers[5]

\journal{arXiv}

\begin{document}
\begin{frontmatter}

\title{A computational approach towards modelling dislocation transmission across phase boundaries}

\author[TUe]{F.~Bormann}
\author[TUe]{R.H.J.~Peerlings}
\author[TUe]{M.G.D.~Geers}
\author[RWTH,MPIE]{B.~Svendsen}

\address[TUe]{Department of Mechanical Engineering, Eindhoven University of Technology, PO Box 513, 5600 MB Eindhoven, The Netherlands}
\address[RWTH]{Material Mechanics, RWTH Aachen University, Schinkelstr. 2, 52062 Aachen, Germany}
\address[MPIE]{Microstructure Physics and Alloy Design, Max-Planck-Institut f\"ur Eisenforschung GmbH, Max-Planck Str. 1, 40237 D\"usseldorf, Germany}

\begin{abstract}
	To study the nanoscopic interaction between edge dislocations and a phase boundary within a two-phase microstructure the effect of the phase contrast on the internal stress field due to the dislocations needs to be taken into account. 
	For this purpose a 2D semi-discrete model is proposed in this paper. It consists of two distinct phases, each with its specific material properties, separated by a fully coherent and non-damaging phase boundary. Each phase is modelled as a continuum enriched with a Peierls--Nabarro (PN) dislocation region, confining dislocation motion to a discrete plane, the glide plane. In this paper, a single glide plane perpendicular to and continuous across the phase boundary is considered. Along the glide plane bulk induced shear tractions are balanced by glide plane shear tractions based on the classical PN model. 
	The model's ability to capture dislocation obstruction at phase boundaries, dislocation pile-ups and dislocation transmission is studied.
	Results show that the phase contrast in material properties (e.g. elastic stiffness, glide plane properties) alone creates a barrier to the motion of dislocations from a soft to a hard phase. 
	The proposed model accounts for the interplay between dislocations, external boundaries and phase boundary and thus represents a suitable tool for studying edge dislocation--phase boundary interaction in two-phase microstructures. 
\end{abstract}

\end{frontmatter}

%
%
\section{Introduction}
Grain and phase boundaries have been shown to strongly influence the ductile behaviour of polycrystalline microstructures, as described, e.g., by the Hall--Petch relation \cite{Hirth1982}. Internal boundaries constitute barriers against dislocation motion, resulting in dislocation obstruction and dislocation pile-ups. Eventually, under sufficiently high shear stress, a variety of events may occur: transmission of a dislocation into the neighbouring grain, dislocation reflection, dislocation nucleation in the neighbouring grain or dislocation migration into the boundary. The aim of this work is to gain a better understanding of the interaction between edge dislocations and phase boundaries related to dislocation obstruction, pile-up and transmission. This involves the non-local interaction between the individual dislocations and also between dislocations and phase boundary. While analytical solutions are limited to simple cases, most non-trivial studies require a numerical treatment. To address the latter, various approaches have been pursued in the literature.
%
%
%
Employing Discrete Dislocation Dynamics (DDD) \cite{Giessen1995, Cleveringa2000} crack growth within single crystals \cite{Giessen2002} and along bi-material interfaces \cite{Needleman2001, ODay2005} has been studied. This method is capable of modelling discrete dislocation activity in domain sizes $>10\mu \mathrm{m}$ and represents an efficient tool to predict corresponding macro-scale material behaviour. However, DDD requires case specific constitutive relations for dislocation--interface interactions (e.g., dislocation transmission). Furthermore, DDD is based on the analytical solution of single dislocations -- often according to the Volterra model -- which is questionable in the vicinity of phase and grain boundaries where the change in dislocation core can play a dominant role.\par
%
%
Molecular dynamics (MD), on the contrary, accounts for the full atomistic complexity of dislocation behaviour. It is thus capable of fully capturing the local interaction between dislocations and the interface. Yet, the utilisation of a proper interatomic potential is essential. This can be a problem within the dislocation core, as pointed out by Das and Gavini \cite{Das2017}. Recently, MD was utilised to study dislocation pile-ups by Wang \cite{Wang2015} in Al with $\Sigma 11$ asymmetrical tilt grain boundaries and by Elzas and Thijsse \cite{Elzas2016} in iron-precipitate microstructures with coherent, semi-coherent and non-coherent interfaces. The complexity of the simulations, however, entails high computational cost. Hence, many simulations are limited to a small number of dislocations. 
%
%
To model larger systems, the quasicontinuum method (QCM) \cite{Miller2002} has been developed. It preserves atomistic accuracy where required and approximates with continuum modelling elsewhere. A particular version of QCM is the coupled atomistic discrete dislocation (CADD) method. It was proposed by Shilkrot et al. \cite{Shilkrot2004} and treats the continuum domain with a DDD framework. Both methods were recently adopted to study dislocation--grain boundary interactions in \cite{Yu2014} (QCM) and \cite{Dewald2007, Dewald2007a, Dewald2011} (CADD). Besides the complex implementation of the pad-region connecting the continuum and the atomistics, the proper choice of the interatomic potential remains a crucial matter.\par
%
%
A full continuum approach that retains the discrete nature of dislocations was introduced by Acharya in \cite{Acharya2001} based on the early work of Kr\"{o}ner \cite{Kroner1958, Kroner1981}, Mura \cite{Mura1963}, Fox \cite{Fox1966} and Willis \cite{Willis1967}. In this theory, dislocations are viewed as sources of elastic incompatibility and are modelled with a dissipative dislocation transport equation. Following De Wit \cite{deWit1970}, Fressengeas \cite{Fressengeas2011} included disclinations to account for rotational defects such as grain boundaries. Recently, Zhang et al. \cite{Zhang2015, Zhang2017} incorporated the effect of lattice periodicity and stacking fault energy to capture dislocation core effects and studied arrays of edge dislocations on one or multiple slip planes in homogeneous and bi-metallic media. The phase boundary is considered as impenetrable and hence introduces an artificial influence on dislocations interacting with the phase boundary.\par
%
%
Another approach was established by Shen and Wang \cite{Shen2004} with a phase-field methodology. The total energy is decomposed in the stacking fault energy of the glide plane, the elastic strain energy following the microelasticity theory of Khachaturyan and Shatalov \cite{Khachaturyan1983} and a gradient energy term for the dislocation core. The Ginzburg--Landau relation describes the evolution of the dislocation structure. Mianroodi et al. \cite{Mianroodi2015, Mianroodi2016} extended this framework to atomistic determination of the gradient term via energy scaling and showed an improved accuracy of the dislocation core structure in comparison with Peierls--Nabarro modelling. In a recent study, Zeng et al. \cite{Zeng2016} studied transmission of a single dislocation across a bi-metallic interface. Considering a lattice mismatch between both phases, coherency stresses and residual dislocations at the phase boundary were incorporated. However, basing the coherency stress purely on the mismatch of material properties, no change in coherency stress due to dislocation interaction is included -- as one would expect from the change of the boundary structure as a result of approaching dislocations. Also, no spreading out of a screw dislocation core into the interface is possible -- as observed by Anderson and Li \cite{Anderson2001} and by Shehadeh et al. \cite{Shehadeh2007}. One of the major drawbacks of this are is computational cost. Hence, domain sizes in the order of $100 nm$ are the largest feasible, resulting in relatively strong image forces arising from the commonly applied periodic boundary conditions.\par
%
%
To investigate the interaction of a single screw dislocation with a slipping and coherent bi-metallic interface, Anderson and Li proposed in \cite{Anderson2001} a simple and computationally efficient approach based on the Peierls--Nabarro (PN) model \cite{Peierls1940, Nabarro1947, Hirth1982}. Following the analytical approach of Paracheco and Mura \cite{Pacheco1969}, they described the screw dislocation by a series of infinitesimally small dislocations with the two-phase stress field by Head \cite{Head1953-Screw} and balanced it against the glide plane tractions of the Frenkel sinusoidal. Shehadeh et al. \cite{Shehadeh2007} extended this model for studying screw dislocation transmission across a non-coherent Cu-Ni interface. The Generalized Stacking Fault (GSF) energy surface \cite{Vitek1968, Joos1994} is utilised for accurate glide plane and phase boundary tractions and additional coherency stresses describe the lattice mismatch between both phases. Similar to the phase field approach by Zeng et al. \cite{Zeng2016}, a change of the coherency stresses due to the local change in phase boundary structure is not accounted for. Additionally, due to its simplicity, an extension towards higher complexities such as phase boundary decohesion or multiple slip systems is not straightforward.\par
%
%
Other approaches based on the PN model concentrated only on studying dislocations in homogeneous media and not on dislocations in two-phase microstructures:
%
%
Schoeck \cite{Schoeck2001, Schoeck2014} studied the dissociation of dislocations and the energy of a dislocation dipole considering the GSF energy surface and using a variational approach with an ansatz for the dislocation profile.
%
%
Rice \cite{Rice1992} suggested a methodology to analyse nucleation of perfect and partial dislocations from stressed crack tips based on the J-integral.
%
%
Xu et al. \cite{Xu1995, Xu1997} extended the variational boundary integral method of Xu and Ortiz \cite{Xu1993} towards the PN dislocation. By modelling the dislocation as a continuous distribution of infinitesimal dislocations \cite{Eshelby1949} they analysed dislocation nucleation from crack tips under mixed mode loading. Later, Xu and Argon \cite{Xu2000} adapted this method towards nucleation of dislocation loops in perfect crystals via a centred circular perturbation and Xu and Zhang \cite{Xu2003} towards dislocation nucleation from crystal surfaces. 
%
%
Xiang et al. \cite{Xiang2008} introduced with the generalised PN model an extension towards 3D, including not only a planar disregistry but also a normal opening of the glide plane. The elastic stress field of a curved dislocation was assessed with Mura's formula. They then studied the core structure of perfect and partial dislocation loops in Cu and Al under different values of applied stress.\par
%
%
In this chapter, the PN model is utilised in 2D plane strain modelling of edge dislocation--phase boundary interaction in two-phase microstructures. Considered is a single glide plane which intersects a fully coherent phase boundary. The glide plane is modelled in accordance with the PN model as an interface that splits the domain into two linear elastic media. Along the glide plane, bulk induced shear tractions balance glide plane shear tractions. The latter originate from the disregistry, the displacement across the glide plane, by a periodic shear traction-disregistry law based on the Frenkel sinusoidal \cite{Frenkel1926}. Individual phases are modelled by adopting phase specific material properties for the elastic medium and the glide plane. The full model is discretised spatially using the finite element method (FEM) and in time by a backward Euler scheme. A highly non-linear problem results, which is solved numerically with an implicit Newton--Raphson algorithm. \par 
Different to previous applications of the PN model, the dislocation induced bulk stresses are not given by an analytical solution, but rather the result of solving, numerically, the governing equations. The model accounts for the interplay between dislocations, external boundary conditions and the phase boundary, while automatically taking the material's properties into account. With only a few parameters it is a rather straightforward model that captures the mechanics of the edge dislocation--phase boundary interaction in an adequate manner. It presents therefore a suitable tool for studying dislocation transmission across phase boundaries. \par
%
%
In this study, a two-phase microstructure is considered, consisting of a soft phase that is flanked by a harder phase. The glide plane is assumed to be perpendicular to the phase boundaries throughout the two phases. On it, at the centre of the soft phase, lies a dislocation source that nucleates dislocation dipoles under sufficiently high shear stress. At the outer boundary of the problem domain a shear deformation is applied which triggers dislocation nucleation and drives the dislocations towards the phase boundary. This configuration is employed to study dislocation obstruction at the phase boundaries, pile-up formation and dislocation transmission into the harder phase.\par
%
%
After the problem statement and model formulation in Section \ref{sect:Model-Intro}, the numerical treatment is detailed in Section \ref{sect:num_implementation}. In Section \ref{sect:Verification} the model is verified by a comparison with analytical solutions and a convergence study is performed. The influence of phase contrast and dislocation pile-ups on dislocation transmission is then studied in Section \ref{sect:Results}. The chapter finishes with conclusions in Section \ref{sect:Conclusion}.

\section{Problem statement and model formulation}
\label{sect:Model-Intro}
\subsection{Problem description}
\begin{figure}[htbp]
	\centering
	\includegraphics[width=1.0\linewidth]{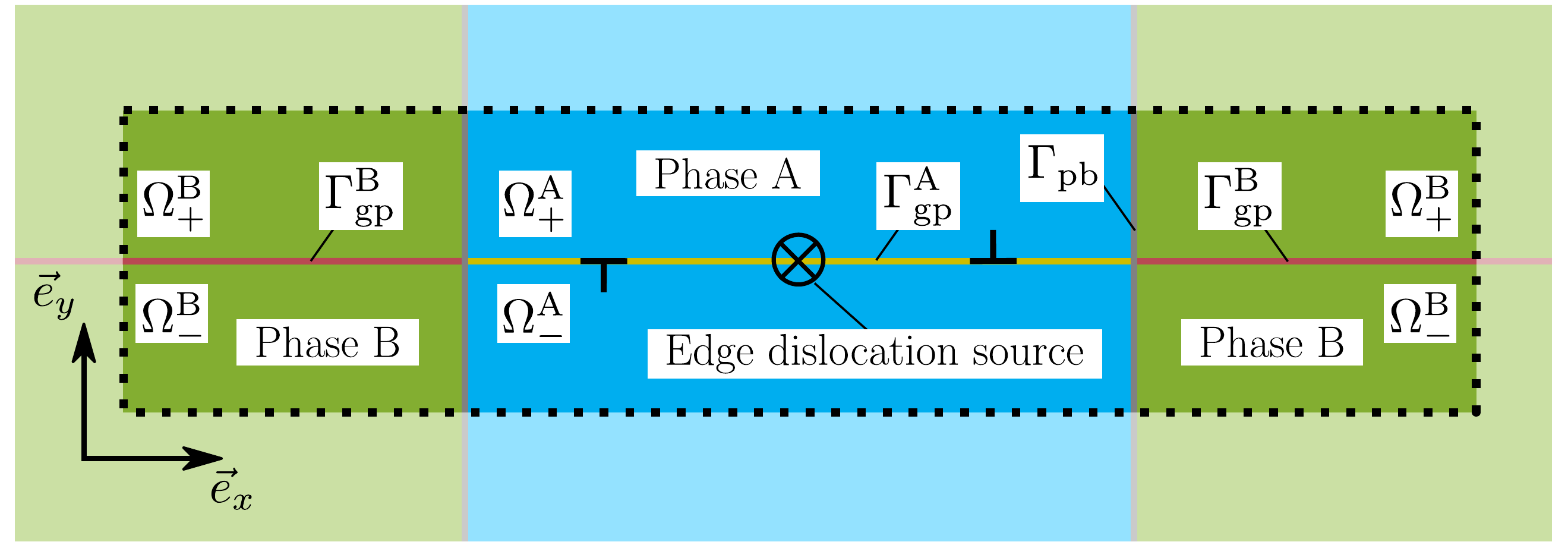}
	\caption{Idealised dislocation motion in a two-phase microstructure with a single glide plane and a centred edge dislocation source.}
	\label{fig:FEM-Model_main_full_simple}
\end{figure}

To study the interaction between edge dislocations and a phase boundary the domain $\Omega$ with its boundary $\partial\Omega$, as sketched in Figure \ref{fig:FEM-Model_main_full_simple}, is considered. Each material point in $\Omega$ is mapped by the position vector $\vec{x}$ in the two-dimensional Euclidean point space $R^2$ with the basis vectors $\vec{e}_x$ and $\vec{e}_y$. Domain $\Omega$ consists of Phase A that is flanked by Phase B, each with distinct material properties. Between both phases lies the shared phase boundary $\Gamma_{\mathrm{pb}}$ which is assumed normal to $\vec{e}_x$
\begin{align}
\Omega &= \Omega^{\mathrm{A}}\cup \Omega^{\mathrm{B}}\\
\partial \Omega &= \left(\partial\Omega^{\mathrm{A}}\setminus\Gamma_{\mathrm{pb}}\right)\cup\left(\partial\Omega^{\mathrm{B}}\setminus\Gamma_{\mathrm{pb}}\right)
\end{align}
where $\Omega^{\mathrm{A}}$ and $\Omega^{\mathrm{B}}$ are the regions occupied by Phases A and B, respectively. In each phase $i\in\{\mathrm{A},\mathrm{B}\}$ lies a single glide plane $\Gamma_{\mathrm{gp}}^{i}$ normal to $\vec{e}_y$ with the slip direction $\vec{e}_x$, splitting $\Omega^{i}$ into $\Omega^{i}_\pm$
\begin{align}
\Omega^{i}&=\Omega^{i}_+\cup\Omega^{i}_-\\
\partial \Omega^{i} &=\left(\partial \Omega^{i}_+\setminus \Gamma_{\mathrm{gp}}^{i}\right)\cup\left(\partial\Omega^{i}_-\setminus\Gamma_{\mathrm{gp}}^{i}\right)
\end{align}
\par
The glide plane is initially defect-free and carries straight edge dislocation lines in the direction $\vec{e}_x \times\vec{e}_y$ only. Dislocations are nucleated from a dislocation source centred in Phase A. It emits a dislocation pair with Burgers vector $\pm b\vec{e}_x$ (a dislocation dipole) when the local resolved shear stress exceeds the critical value $\tau_{\mathrm{nuc}}$, representing the activation stress of a Frank--Read source. The resolved shear stress results from a shear deformation applied on the remote boundary $\partial\Omega$ and the dislocation induced internal stress field. A freshly nucleated dipole starts moving under the local Peach--Koehler force towards the phase boundary, where it will interact with this interface. \par
For the present study, the glide plane $\Gamma_{\mathrm{gp}}$ is taken perpendicular to and continuous across a fully coherent phase boundary $\Gamma_{\mathrm{pb}}$. Interfacial decohesion is inhibited. Only dislocation motion from a soft Phase A to a stiff Phase B is considered. Accordingly, two possible dislocation--phase boundary interactions may exist: (1) dislocation obstruction and (2) dislocation transmission into Phase B. Evidently, dislocation obstruction may lead to the formation of a dislocation pile-up in which dislocations exert repelling forces on each other.
%
%
\subsection{Governing equations}
Under the absence of body forces the equilibrium of domain $\Omega$ is governed by the momentum balance in $\Omega_\pm^{i}$, along $\Gamma_{\mathrm{gp}}^{i}$ and along $\Gamma_{\mathrm{pb}}$:
\begin{align}
\label{eq:balance-bulk}
\mathrm{div}\; \boldsymbol{\sigma}^{i} &= \vec{0} &&\mathrm{in}\enspace \Omega_\pm^{i}\\
\label{eq:balance-gp}
\vec{T}_{\mathrm{gp},-}^{i} &=\vec{T}_{\mathrm{gp},+}^{i} &&\mathrm{on}\enspace \Gamma_{\mathrm{gp}}^{i}\\
\label{eq:pb-traction}
\vec{T}_{\mathrm{pb}}^{\mathrm{A}}&=\vec{T}_{\mathrm{pb}}^{\mathrm{B}} &&\mathrm{on}\enspace \Gamma_{\mathrm{pb}}
\end{align}
where $\boldsymbol{\sigma}^{i}$ is the stress tensor in $\Omega_\pm^{i}$, $\vec{T}^{i}_{\mathrm{gp},\pm} = \boldsymbol{\sigma}^{i}_\pm\cdot\vec{e}_y$ the shear tractions acting on $\partial\Omega^{i}_\pm\cap\Gamma^{i}_{\mathrm{gp}}$ and $\vec{T}_{\mathrm{pb}}^{i} = \boldsymbol{\sigma}^{i}\cdot\vec{e}_x$ the shear tractions acting on $\partial\Omega^{i}\cap\Gamma_{\mathrm{pb}}$. The perfectly bonded and non-damaging phase boundary $\Gamma_{\mathrm{pb}}$ is enforced by the displacement continuity
\begin{equation}
\label{eq:pb-displacement}
\vec{u}^{\mathrm{A}} = \vec{u}^{\mathrm{B}}\quad \mathrm{on}\enspace \Gamma_{\mathrm{pb}}
\end{equation}
\par
Following the concept of the classical Peierls--Nabarro (PN) model \cite{Peierls1940, Nabarro1947, Hirth1982}, linear elastic material behaviour under a plane strain condition is adopted in each subdomain $\Omega^{i}_\pm$ at time $t$:
\begin{align}
\label{eq:el-stress-strain}
\boldsymbol{\sigma}^{i} &= {}^4\boldsymbol{C}^{i}:\boldsymbol{\varepsilon}\\
\label{eq:strain-disp}
\boldsymbol{\varepsilon} &= \frac{1}{2}\left(\vec{\nabla} \vec{u} + \left(\vec{\nabla} \vec{u}\right)^T\right)
\end{align}
where $^4\boldsymbol{C}^{i}$ is the fourth-order elasticity tensor associated with Phase $i$. Along the glide plane $\Gamma_{\mathrm{gp}}^{i}$ a large displacement jump $\vec{\Delta}$ is allowed for that originates from dislocation glide:
\begin{equation}
\label{eq:disregistry}
\vec{\Delta} = \left[\left[\vec{u}\,\right]\right]=\vec{u}_+-\vec{u}_-
\end{equation}
with
\begin{equation}
\label{eq:disregistry-vector}
\vec{\Delta} = \Delta_t \vec{e}_x
\end{equation}
and a zero normal component $\Delta_n$. Associated with $\Delta_t$, in the following denoted disregistry $\Delta$, is a glide plane shear traction $T^{i}(\Delta)$ which defines the vector $\vec{T}^{i} = \vec{T}^{i}_{\mathrm{gp},\pm}$:
\begin{equation}
\label{eq:traction-pm}
\vec{T}^{i} = T^{i}(\Delta)\,\vec{e}_x
\end{equation}
In this chapter a periodic shear traction-disregistry law is employed as follows
\begin{equation}
\label{eq:PN-num}
T^{i}(\Delta) = \widehat{T}^{i}\,\sin\left(\frac{2\pi \Delta}{b^{i}}\right)+\eta^{i}\dot{\Delta}
\end{equation}
with the traction amplitude $\widehat{T}^{i}$ and the damping coefficient $\eta^{i}$. The first term in Eq.~\eqref{eq:PN-num} is the Frenkel sinusoidal function \cite{Frenkel1926} and represents the periodic restoring force acting between the initially aligned atoms above and below the glide plane. The second term in Eq. \eqref{eq:PN-num}, with the damping coefficient $\eta^{i}$, is an addition to the classical approach and introduces a time dependency in the solution $\vec{u}(\vec{x},t)$. Using this term, the dynamic evolution of the dislocations is regularised and numerical stability is guaranteed. As shown in Appendix \ref{Appendix:Damping}, $\eta$ can be related to the drag coefficient commonly used in DDD. The choice of a Frenkel sinusoidal function follows the classical PN model. As shown by Rice \cite{Rice1992}, $\Delta$ not only includes the inelastic, but also the elastic disregistry between the two adjacent atomic layers. Although this implies that the elastic response of that layer is accounted for twice, qualitative studies remain possible and valuable. A physically more realistic shear traction-disregistry relation is based on the Generalized Stacking Fault energy (GSFE) \cite{Joos1994}. The GSFE can be obtained from atomistic calculations \cite{Vitek1968, Lejcek1972, Kroupa1972} or from the density functional theory \cite{Duesbery1990, Kaxiras1993}.\par
In the classical PN model a single static dislocation in a homogeneous and infinite medium is considered. The analytical solution for the disregistry profile can then be obtained as \cite{Hirth1982}
\begin{equation}
\label{eq:disreg-analytical}
\Delta(x) = -\frac{b}{\pi}\tan^{-1}\left(\frac{x-x_1}{\zeta}\right)+\frac{b}{2},\
\end{equation}
where $x_1$ denotes the dislocation position and $\zeta$ the dislocation core half-width (in an isotropic crystal \cite{Joos1994})
\begin{equation}
\label{eq:dislocation-width}
\zeta = \frac{\mu b}{4\pi\left(1-\nu\right)\widehat{T}}
\end{equation} 
with the shear modulus $\mu$ and Poisson's ratio $\nu$. For a two-phase microstructure no analytical solution is available and a numerical solution strategy is thus required.
\subsection{Boundary conditions}
\begin{figure}[htbp]
	\centering
	\includegraphics[width=0.6\linewidth]{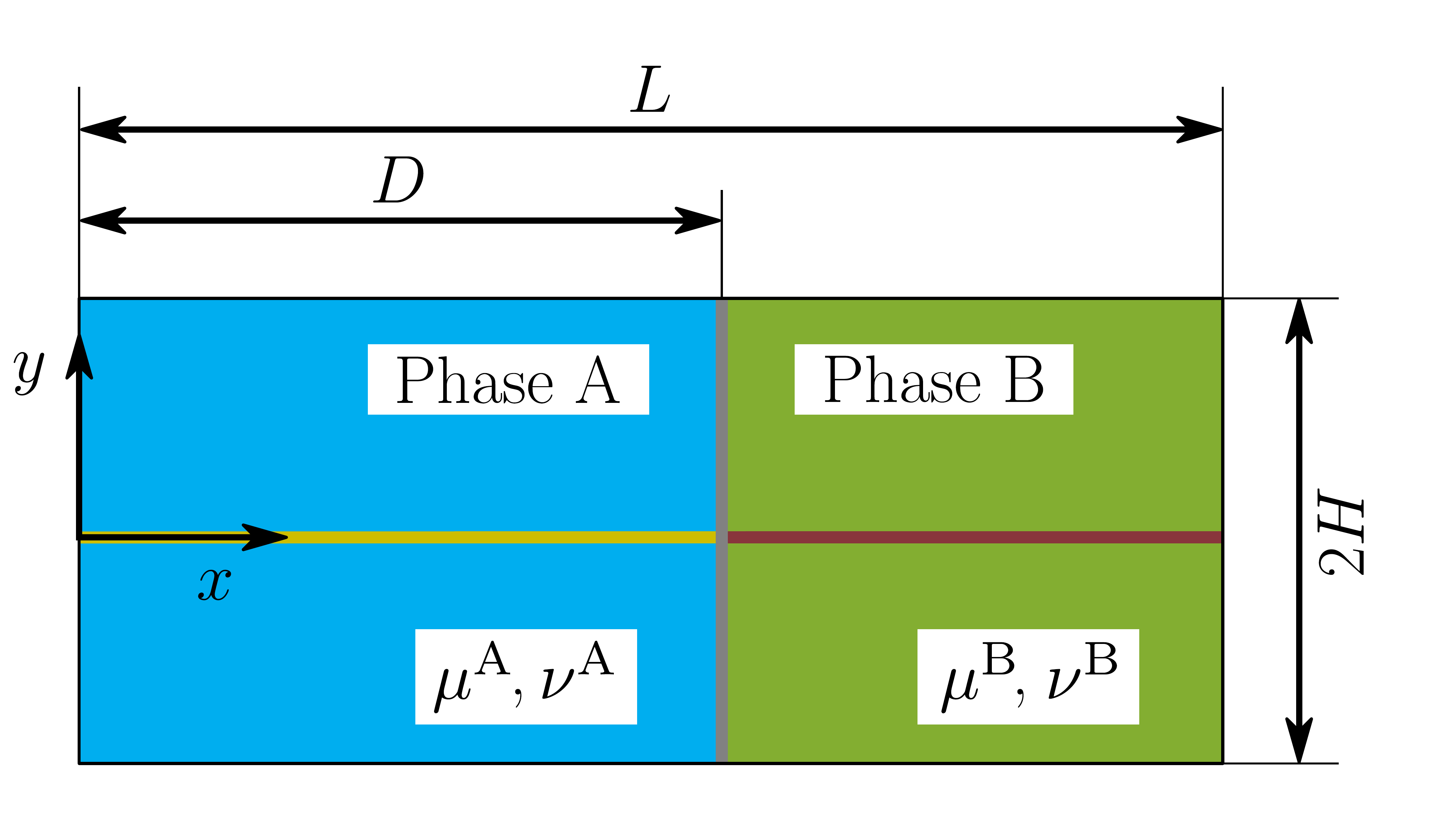}
	\caption{Geometrical two-phase model for edge dislocation--phase boundary interaction with point-symmetry at $x=0$, $y=0$.}
	\label{fig:FEM-Model_main_half}
\end{figure}

Considering the model in Figure \ref{fig:FEM-Model_main_full_simple}, it is apparent that the solution of the problem has the symmetry property $\vec{u}(-\vec{x})=-\vec{u}(\vec{x})$ with respect to the dislocation source ($\vec{x} = \vec{0}$ with respect). It is hence possible to reduce the model domain to the half-model presented in Figure \ref{fig:FEM-Model_main_half} with model height $2H$, length $L$ and length $L_{\mathrm{A}}$ of Phase A. The symmetry boundary condition reads
\begin{equation}
\label{eq:point-symmetry}
\vec{u}(y) = -\vec{u}(-y)\quad\mathrm{on}\:\Gamma_{\mathrm{ps}}= \{(0,y)| -H<y<H\}
\end{equation}
The shear deformation is imposed on $\partial\Omega\setminus\Gamma_{\mathrm{ps}}$ in form of displacements that are consistent with a purely linear elastic model response -- neglecting the glide plane -- which would result under an applied shear load $\tau$. Correspondingly, $\tau$ is used to parametrise the Dirichlet boundary conditions
\begin{align}
\label{eq:boundary_value1}
u_y&=\frac{\tau}{\mu^{\mathrm{A}}}x && \quad\mathrm{on}\:\partial\Omega^{\mathrm{A}}\setminus\left(\Gamma_{\mathrm{pb}}\cup\Gamma_{\mathrm{ps}}\right)\\
\label{eq:boundary_value2}
u_y&=\frac{\tau}{\mu^{\mathrm{B}}}(x-L_{\mathrm{A}})+\frac{\tau}{\mu^{\mathrm{A}}}L_{\mathrm{A}} && \quad\mathrm{on}\:\partial\Omega^{\mathrm{B}}\setminus\Gamma_{\mathrm{pb}}\\
\label{eq:boundary_value3}
u_x&= 0 && \quad\mathrm{on}\:\partial\Omega\setminus\Gamma_{\mathrm{ps}}
\end{align}
%
\section{Numerical implementation}
\label{sect:num_implementation}
\subsection{Finite element discretisation}
The weak form of the balance of momentum \eqref{eq:balance-bulk} and \eqref{eq:balance-gp} with the constitutive relations \eqref{eq:el-stress-strain}, \eqref{eq:strain-disp} and \eqref{eq:PN-num} and the phase boundary constraints \eqref{eq:pb-displacement} and \eqref{eq:pb-traction} can be derived from the Generalized Clapeyron’s formula \cite[Chapter 5.14]{Asaro2006} as
\begin{equation}
\label{eq:weak-form-lin}
\int_{\Omega\setminus\Gamma_{\mathrm{gp}}}\left(\vec{\nabla}\vec{u}_*\right):{}^4\boldsymbol{C}:\left(\vec{\nabla}\vec{u}\right)\,\mathrm{d}\Omega + \int_{\Gamma_{\mathrm{gp}}} \vec{\Delta}_* \cdot\vec{T}\,\mathrm{d}\Gamma=0
\end{equation}
Here, $\vec{u}_*$ is the test displacement and $\vec{\Delta}_*=\left[\left[\vec{u}_*\right]\right]$ the test disregistry. In order to solve Eq. \eqref{eq:weak-form-lin} numerically for the displacement field $\vec{u}(\vec{x},t)$ a discretisation in time and space is required. For the temporal discretisation the backward Euler method is adopted. Let $\vec{u}(\vec{x}, t)$ be a known solution. An approximation of the velocity $\dot{\vec{u}}$ is given by
\begin{equation}
\dot{\vec{u}} \approx \frac{\vec{u}\left(\vec{x},t+\theta \right)-\vec{u}\left(\vec{x},t\right)}{\theta} = \frac{\vec{u}-\vec{u}^t}{\theta}
\end{equation}
where $\theta$ is the time step. Hence, the shear traction-disregistry law \eqref{eq:PN-num} reads in discretised form
\begin{equation}
\label{eq:PN-disc}
T^{i} = \widehat{T}^{i}\sin\left(\frac{2\pi\Delta}{b^{i}}\right)+\eta^{i}\frac{\Delta-\Delta^t}{\theta}
\end{equation}
This leads to a non-linearity of the weak form which is solved implicitly with a Newton-Raphson scheme at each time $t+\theta$:  
\begin{align}
\label{eq:weak-form-nl}
&\int_{\Omega\setminus\Gamma_{\mathrm{gp}}}\left(\vec{\nabla}\vec{u}_*\right):{}^4\boldsymbol{C}:\left(\vec{\nabla}\delta\vec{u}\right)\,\mathrm{d}\Omega + \int_{\Gamma_{\mathrm{gp}}} \vec{\Delta}_* \cdot\delta\vec{T}\,\mathrm{d}\Gamma=\\\nonumber
-&\int_{\Omega\setminus\Gamma_{\mathrm{gp}}}\left(\vec{\nabla}\vec{u}_*\right):{}^4\boldsymbol{C}:\left(\vec{\nabla}\vec{u}\right)\,\mathrm{d}\Omega - \int_{\Gamma_{\mathrm{gp}}} \vec{\Delta}_* \cdot\vec{T}\,\mathrm{d}\Gamma
\end{align}
where $\vec{u}$ and $\vec{T}$ denote the computed state variables at the previous iteration $k$ and $\delta \vec{u}$ is the iterative correction to be solved for. $\delta\vec{T}$ is related to $\delta u$ via Eq. \eqref{eq:disregistry} and 
\begin{equation}
\delta\vec{T} = \left.\frac{\partial \vec{T}}{\partial \vec{\Delta}}\right|_{\Delta = \Delta_k}\cdot\delta\vec{\Delta}=\boldsymbol{M}\cdot\delta\vec{\Delta}
\end{equation}
with 
\begin{equation}
\boldsymbol{M} = M\;\vec{e}_x\vec{e}_x
\end{equation}
and, from Eq. \eqref{eq:PN-disc},
\begin{equation}
M=\widehat{T}^{i}\frac{2\pi}{b^{i}}\cos\left(\frac{2\pi\Delta_k}{b^{i}}\right)+\frac{\eta^{i}}{\theta}
\end{equation}
For the spatial discretisation use is made of the FEM with Lagrange shape functions. Adopting the Voigt notation, Eq. \eqref{eq:weak-form-nl} can be rewritten as
\begin{equation}
\label{eq:weak-form-disc-nl}
\left(\int_{\Omega\setminus\Gamma_{\mathrm{gp}}}\underline{B}^T\underline{C}\,\underline{B}\,\mathrm{d}\Omega
+ \int_{\Gamma_{\mathrm{gp}}} \underline{P}^T \utilde{\bar{N}}^T \underline{M} \utilde{\bar{N}}\underline{P} \,\mathrm{d}\Gamma\right)\cdot\delta\utilde{u}=-\utilde{f}^{\:\mathrm{int}}
\end{equation}
with
\begin{equation}
\label{eq:f-int}\textbf{}
\utilde{f}^{\:\mathrm{int}} = \left(\int_{\Omega\setminus\Gamma_{\mathrm{gp}}}\underline{B}^T\underline{C}\,\underline{B}\,\mathrm{d}\Omega\right)\utilde{u} 
+ \int_{\Gamma_{\mathrm{gp}}} \underline{P}^T \utilde{\bar{N}}^T \utilde{T}\,\mathrm{d}\Gamma
\end{equation}
Here, $\utilde{N}$ and $\utilde{\bar{N}}$ are the shape functions for bulk and glide plane, respectively, $\underline{B}$ contains the spatial derivatives of the shape functions in $\utilde{N}$, $\underline{C}$ is the Voigt elasticity matrix, $\underline{M}$ is the glide plane tangent stiffness, $\utilde{u}$ are the nodal displacements and $\utilde{T}$ are the nodal glide plane tractions. $\underline{P}$ is a matrix such that in accordance with Eq. \eqref{eq:disregistry}
\begin{equation}
\utilde{\Delta}=\underline{P}\,\utilde{u}
\end{equation}
holds, e.g., for the linear quadrilateral interface element
\begin{equation}
\begin{bmatrix}
\Delta_t^{1-4}\\
\Delta_t^{2-3}
\end{bmatrix}
=
\begin{bmatrix}
-1	&	0	&	0	&	1\\
0	&	-1	&	1	&	0	
\end{bmatrix}
\begin{bmatrix}
u_t^1\\
u_t^2\\
u_t^3\\
u_t^4\\
\end{bmatrix}
\end{equation}
with nodes 1 and 2 of $\Omega_-$ initially coincident with nodes 4 and 3 of $\Omega_+$. Tie constraints for the description of the point symmetry \eqref{eq:point-symmetry} and the glide plane constraint $\Delta_n=0$ are adopted in a standard FEM manner.\par
Each domain $\Omega_\pm^{i}$ is discretised by bilinear quadrilateral elements with four Gauss integration points. For the glide plane linear interface elements of zero thickness with two Gauss integration points are used.\par
%
\subsection{Dislocation nucleation}
\label{sect:disl_nuc}
As stated in Section \ref{sect:Model-Intro}, a dislocation source is used to nucleate dislocation dipoles each time the resolved shear stress $\tau^{\mathrm{res}}$ at the position of the source exceeds the critical value $\tau_{\mathrm{nuc}}$. The numerical implementation entails: (1) checking the condition $\tau^{\mathrm{res}}\ge\tau_{\mathrm{nuc}}$; (2) introduction of the positive part of the dipole at a certain position $x$ where the dislocation (dipole) close to its stable equilibrium position. Latter is motivated by the computational expense that is required to propagate a dislocation (dipole) from its nucleation position towards equilibrium. Focusing on the process of dislocation transmission and not dislocation motion towards the pile-up, a nucleation close to the static equilibrium is desirable.\par
For the nucleation criterion, the (converged) mean resolved shear stress $\tau_{\mathrm{m}}^{\mathrm{res}}$ is computed in the vicinity of the dislocation source using Eq. \eqref{eq:PN-disc}:
\begin{equation}
\tau_{\mathrm{m}}^{\mathrm{res}}=\frac{1}{L_0}\int_{0}^{L_0} T^{\mathrm{A}}\,\mathrm{d}x
\end{equation}
The disregistry rate $(\Delta-\Delta^t)/\delta t$ plays only a negligible role since either a potential dislocation is very close and the sinusoidal term is highly negative (no dislocation nucleation) or the dislocation is distant and the disregistry rate insignificant (potential dislocation nucleation). For the point-symmetric model the interval $0\le x\le L_0$ is used to average the resolved shear stress. For the distance $L_0$, use is made of the DDD equivalent for a newly nucleated dipole as formulated by Van der Giessen and Needleman \cite{Giessen1995}:
\begin{equation}
\label{eq:min-nucl-dist}
L_0=\frac{1}{2}\frac{\mu^{\mathrm{A}}b^{\mathrm{A}}}{2\pi(1-\nu)\tau_{\mathrm{nuc}}}
\end{equation}\par
Once the nucleation criterion is fulfilled, a positive dislocation is introduced (and hence due to the point-symmetry also the dipole) by solving an intermediate BVP with new Dirichlet boundary conditions. At the outer boundary one requires
\begin{equation}
\label{eq:dislocation-intro-bc}
\delta \utilde{u} = 0 \quad\mathrm{on}\enspace\partial\Omega\setminus\Gamma_{PS}
\end{equation}
Additionally, at the first iteration, the disregistry profile of the to be nucleated dislocation is imposed by the analytical solution \eqref{eq:disreg-analytical}:
\begin{equation}
\delta \utilde{u}_{x,\pm} = \pm\frac{b^{\mathrm{A}}}{2}\left(-\frac{1}{\pi}\tan^{-1}\left(\frac{x-L_{\mathrm{nuc}}}{\zeta^{\mathrm{A}}}\right)+\frac{1}{2}\right)\quad\mathrm{on}\enspace\Gamma_{\mathrm{gp}}
\end{equation}
where $\delta{\utilde{u}}_{x,\pm}$ are the displacements belonging to $\partial\Omega_\pm \cap \Gamma_{\mathrm{gp}}$, and $L_{\mathrm{nuc}}$ the nucleation position of the new dislocation. During the subsequent iterations $\delta \utilde{u}_{x,\pm} = 0$ applies. After solving the intermediate BVP, the simulation is continued with the previously defined Dirichlet boundary conditions -- and an unconstrained glide plane. \par
To find a suitable nucleation position $L_{\mathrm{nuc}}$, the position $x$ is sought at which the current glide plane shear tractions $T(x)$ and the shear stresses $\sigma_{xy}^\pm(x)$ of the to be nucleated positive/negative dislocation, situated at $\pm x$, are in equilibrium:%
\begin{equation}
\label{eq:disl-intro-pos}
T(x)+\sigma_{xy}^+(x)+\sigma_{xy}^-(x) = 0
\end{equation}
Neglecting the influence of the left boundary, a good approximation for $\sigma_{xy}^\pm(x)$ is given by Head \cite[Section 10]{Head1953-Edge} who derived the stress field of a dislocation in a bi-metallic medium.
Mathematically, Eq. \eqref{eq:disl-intro-pos} may have $j+1$ stable equilibrium positions and one unstable equilibrium position in $L_0\le x\le L_{\mathrm{A}}$. However, for $j>0$, only one stable equilibrium position lies between $L_0$ and the pile-up ($\min{x_{j}}$). Hence, an upper bound for the dislocation nucleation position is introduced as
\begin{equation}
\label{eq:nucl-max}
L_{\mathrm{max}} = 
\begin{cases}
L_{\mathrm{A}}\qquad & \mathrm{for}\enspace j=0\\
\min(x_{j})\qquad & \mathrm{for} \enspace j>0
\end{cases}
\end{equation} 
The unstable equilibrium position characterises the point at which a small perturbation leads either to annihilation of the dipole or to dislocation motion towards the phase boundary into the stable equilibrium position. Thus, the nucleation position is chosen as the position $x\in[L_0,L_{\mathrm{max}}]$ where Eq. \eqref{eq:disl-intro-pos} holds and that is the closest to the phase boundary ($x=L_{\mathrm{A}}$).
%
\section{Verification and convergence study}
\label{sect:Verification}
\subsection{Single dislocation in a homogeneous medium}
The verification of the implemented PN continuum model is accompanied by a mesh convergence study in a single dislocation framework. The numerically obtained stress field of the dislocation in a homogeneous material is compared with the analytical solution presented in \cite{Hirth1982}, i.e., a true verification step. Considered is a model of dimensions $L\times 2 H = 100 b \times 100 b$; uniform meshes are employed, with element sizes of $h=b/8,\,b/4,\,b/2,\,b,\,2b$. For the linear elastic behaviour of the single Phase A considered, a shear modulus of $\mu^{\mathrm{A}} = 80.77 \,\mathrm{GPa}$ and a Poisson's ratio of $\nu^{\mathrm{A}} = 0.3$ are adopted. The glide plane amplitude is chosen in alignment with the small strain limit of linear elasticity as $\widehat{T}^{\mathrm{A}} = \mu^{\mathrm{A}} b^{\mathrm{A}}/2\pi d^{\mathrm{A}}$, where $d^{\mathrm{A}}$ is the interplanar spacing \cite{Hirth1982}. For simplicity $b^{\mathrm{A}}$ and $d^{\mathrm{A}}$ are assumed to be equal such that $\widehat{T}^{\mathrm{A}}=\mu^{\mathrm{A}}/2\pi$. The Burgers vector is $b=b^{\mathrm{A}}=0.25 \,\mathrm{nm}$. Following Appendix \ref{Appendix:Damping}, the viscosity is defined by the relationship $\eta^{\mathrm{A}}=B/b$ in terms of the Drag coefficient $B = 10^{-4}\,\mathrm{Pa\,s}$. The critical nucleation stress is set to $\tau_{\mathrm{nuc}}=0.048\,\widehat{T}^{\mathrm{A}}$.\par
\begin{figure}[htbp]
	\centering
	\includegraphics[width=1.0\linewidth]{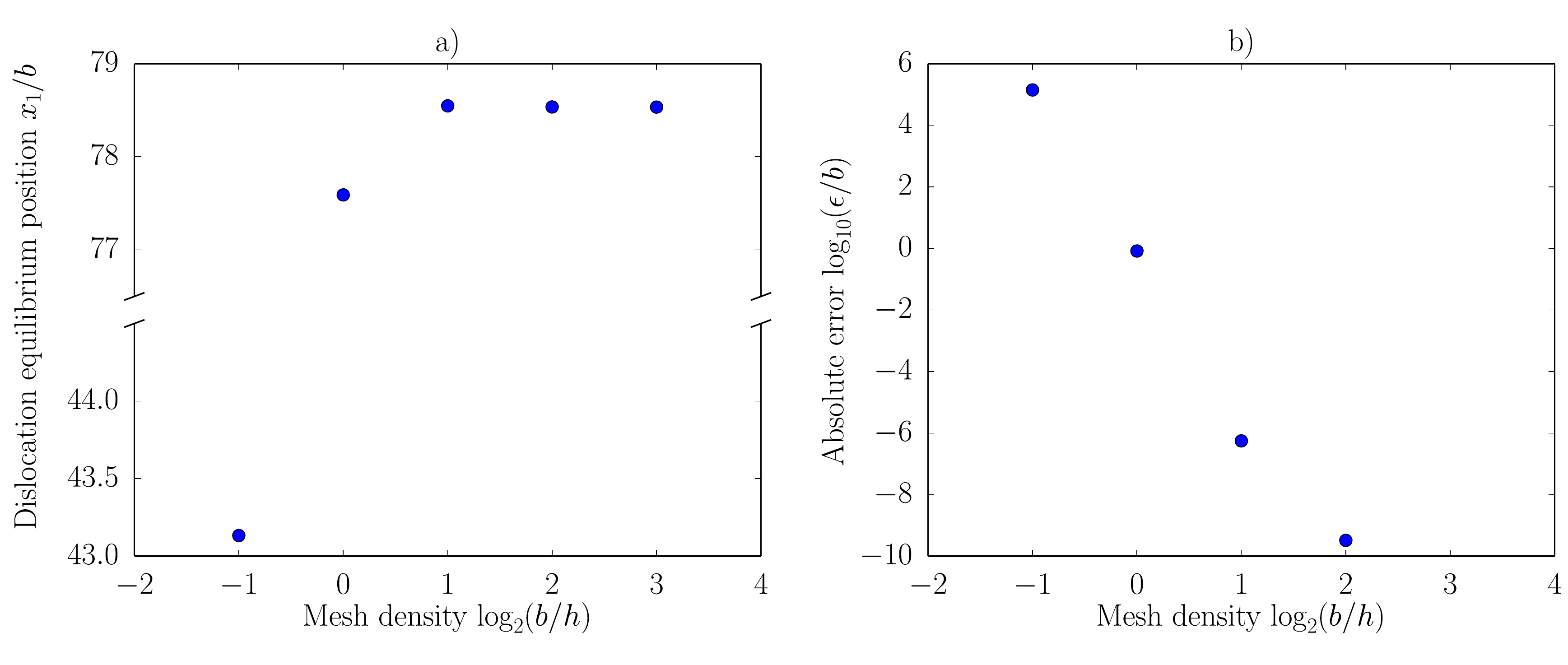}
	\caption{Influence of the discretisation on the dislocation equilibrium position.}
	\label{fig:Mesh-dependency-disl-pos}
\end{figure}

With the application of a constant shear load of $\tau=0.054\,\widehat{T}^{\mathrm{A}}$ dislocation nucleation is triggered, enforced here at $L_{\mathrm{nuc}}\approx 40.5b$. Subsequently, the dislocation moves in accordance with the BVP towards the rigid surface at $x=L$ and settles in its equilibrium position $x_1$. Here, the dislocation position $x_1$ is defined as the coordinate where $\Delta = b/2$. Figure \ref{fig:Mesh-dependency-disl-pos}a shows the final dislocation position $x_1$ as a function of the element size $h$. The convergence behaviour is presented in Figure \ref{fig:Mesh-dependency-disl-pos}b where the absolute error with respect to the finest discretisation considered, $\epsilon=\left|x_1(h=b/8)-x_1(h)\right|$, is plotted against the mesh size $h$. A strong influence of the discretisation on the final dislocation position is evident for coarse discretisations. Mesh sizes of $h\ge b$ lead to an absolute error of $\epsilon\approx b$ and above. For element sizes $h\le b/2$, however, the dislocation position quickly converges, resulting in an accuracy on the order of $10^{-6}\,b$ or better. A (too) coarse discretisation is unable to represent the gradient of the disregistry and leads hence to a large deviation of the dislocation position. \par
\begin{figure}[htbp]
	\centering
	\includegraphics[width=1.0\linewidth]{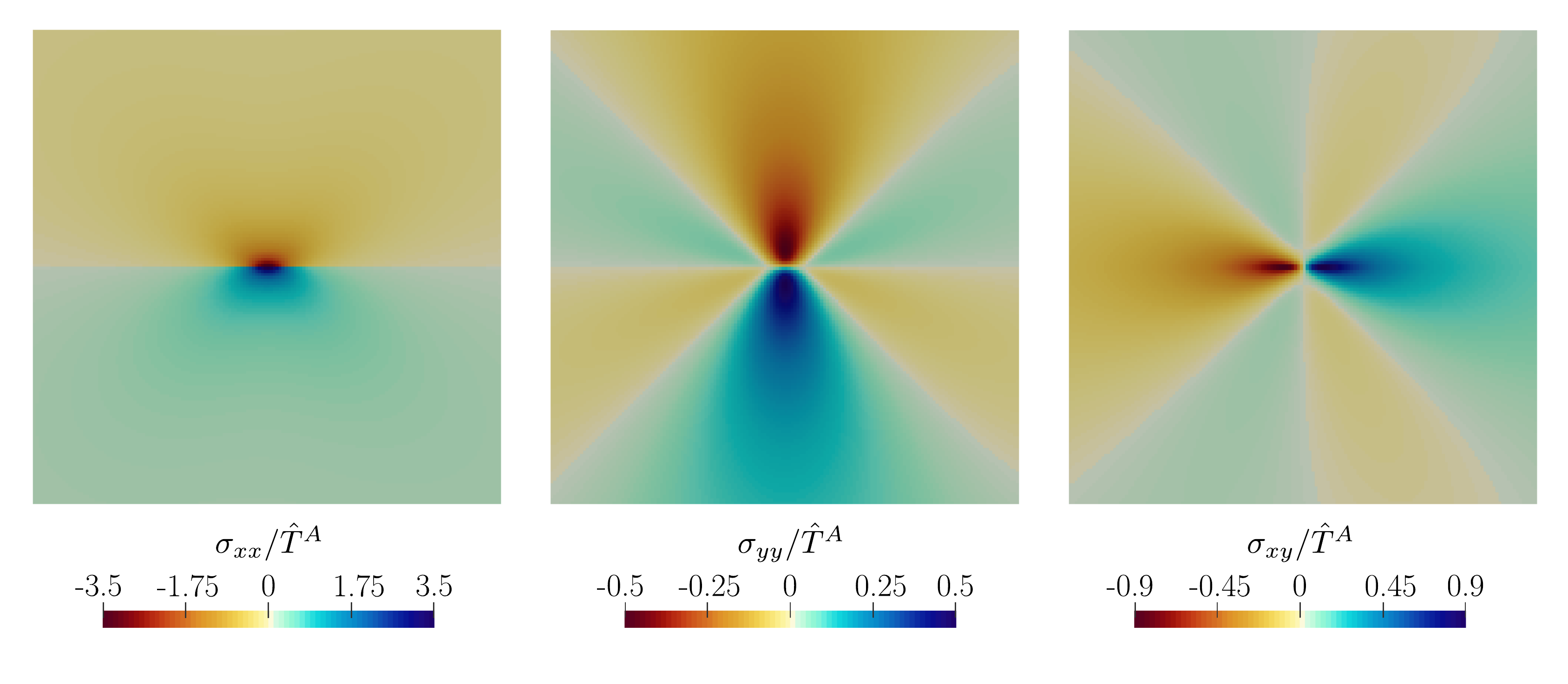}
	\caption{2D stress field of $h=b/8$ at equilibrium.}
	\label{fig:2D-Stress_field}
\end{figure}

\begin{figure}[htbp]
	\centering
	\includegraphics[width=1.0\linewidth]{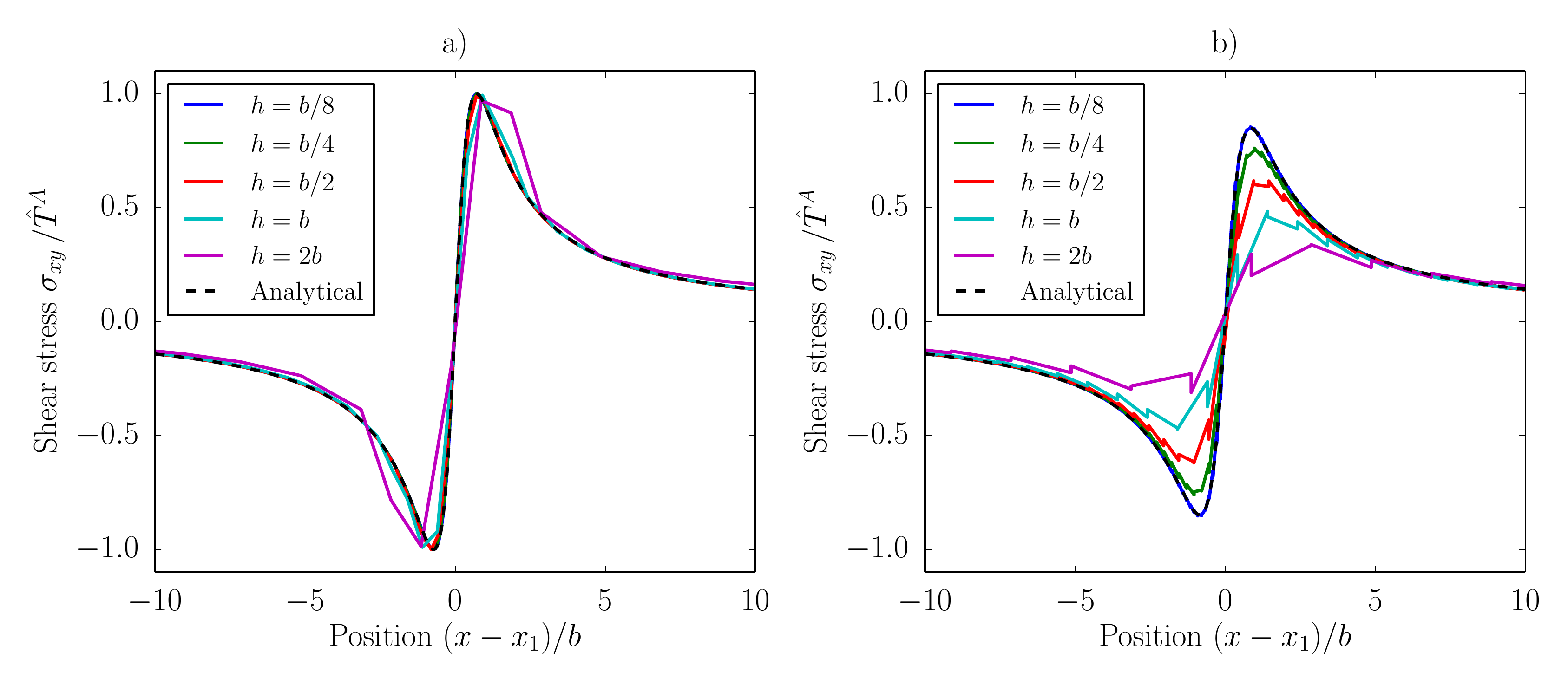}
	\caption{Shear stress $\sigma_{xy}(x-x_1,y)$ at (a) $y=0$ and (b) $y=b/16$ for different discretisation sizes, along with the analytical solution.}
	\label{fig:Shear-stress-line_plot}
\end{figure}

Studying the dislocation induced stress field, one recovers the well-known 2D stress field, as shown in Figure \ref{fig:2D-Stress_field} for $h=b/8$. 
To visualise the discretisation influence on the stress field representative line plots are extracted. In Figure \ref{fig:Shear-stress-line_plot}a and \ref{fig:Shear-stress-line_plot}b the shear stress $\sigma_{xy}$ is plotted as a function of the position $x-x_1$ along the $x$-axis at $y=0$ and $y=b/16$, respectively. In both figures, the analytical solution for the infinite domain is added as a reference. 
For $\sigma_{xy}(x,y=0)$, the glide plane shear traction $T$ provides a rather accurate measure. The shear stress is captured adequately for mesh sizes of $h\le b/2$. For the discretisation $h=b$, small discrepancies emerge, while for $h=2b$ the results become too inaccurate. This is in alignment with the convergence study of the dislocation position and shows that the proper quantification of the shear tractions plays a major role for accuracy of the dislocation evolution. The larger deviations that are observed for $\sigma_{xy}(x,y=b/16)$ (Figure \ref{fig:Shear-stress-line_plot}b) illustrate the significant impact of the spatial discretisation on the shear stress $\sigma_{xy}$. Besides a vertical offset, also stress jumps across the elements arise. Both offset and stress jumps are reduced by refining the mesh size.\par
\begin{figure}[htbp]
	\centering
	\includegraphics[width=1.0\linewidth]{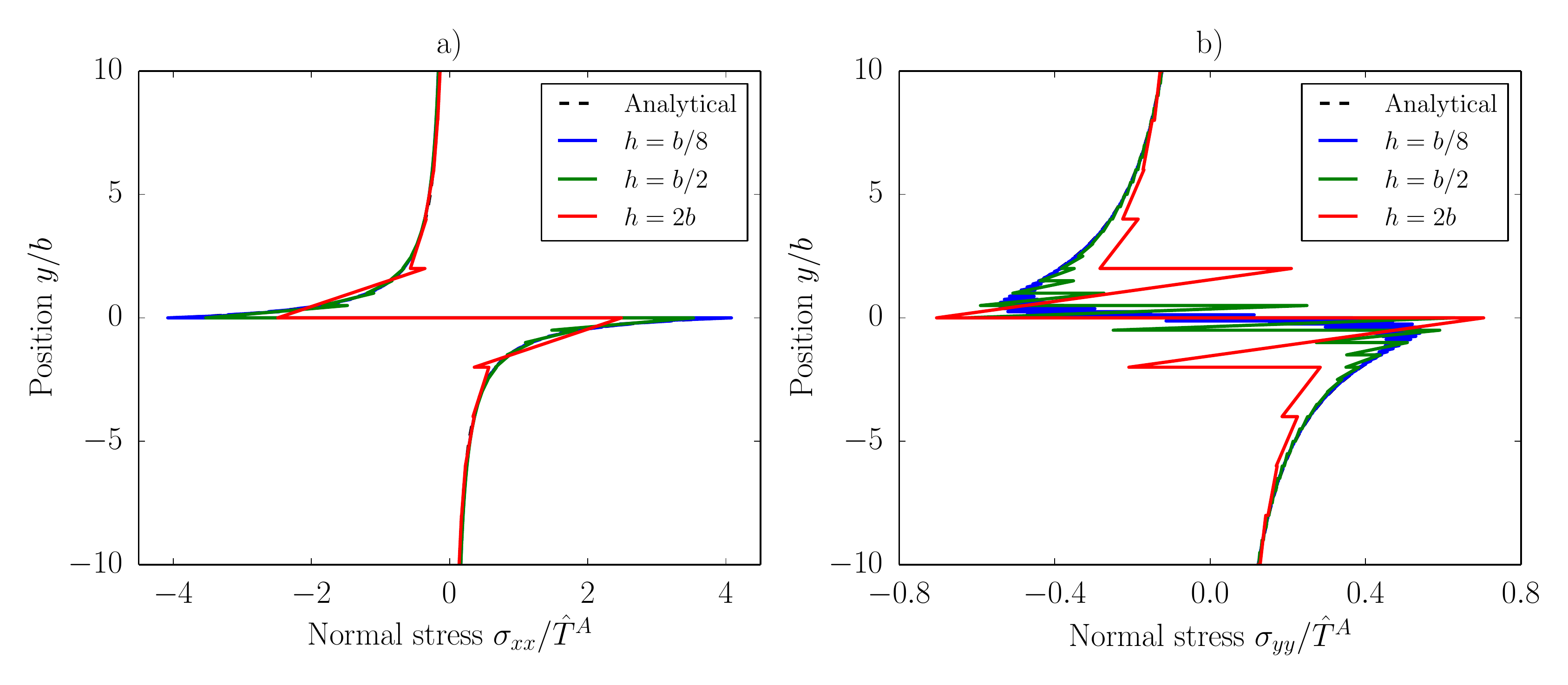}
	\caption{Normal stresses (a) $\sigma_{xx}(x=x_1, y)$ and (b) $\sigma_{yy}(x=x_1, y)$ for different discretisation sizes along with the analytical solution.}
	\label{fig:Normal-stress-line_plot}
\end{figure}

Similar stress jumps are observed for $\sigma_{xx}(x=x_1,y)$ and $\sigma_{yy}(x=x_1,y)$, with significant irregularities in the latter one, as visualised in Figures \ref{fig:Normal-stress-line_plot}a and \ref{fig:Normal-stress-line_plot}b, respectively. For the sake of clarity only the stresses for $h=b/8,\,b/2,\,2$ are presented. Note that $\sigma_{xx}$ exhibits smaller fluctuations due to its smoother stress gradient in comparison to $\sigma_{xy}$ and $\sigma_{yy}$. Nevertheless, the solution converges towards the analytical solution for increasingly fine meshes. Different to the glide plane traction $T$ and the related dislocation position $x_1$ where an element size of $h \le b/2$ suffices, a rather fine mesh with $h \le b/8$ is required for a high accuracy in the stress response.\par
%

%
\subsection{Two-phase microstructure and dislocation transmission}
\label{sect:Transmission-Verification}
For the study of dislocation obstruction and dislocation transmission, a two-phase microstructure is considered with the grain size of Phase A $2L_{\mathrm{A}}=L$. The properties of Phase A are identical to those introduced in Section \ref{sect:Verification}. For Phase B the following material properties are adopted: $\mu^{\mathrm{B}} = 7\mu^{\mathrm{A}}/6$, $\widehat{T}^{\mathrm{B}}=7\,\widehat{T}^{\mathrm{A}}/6$, $\nu^{\mathrm{B}}=\nu^{\mathrm{A}}$, $b^{\mathrm{B}}=b$ and $\eta^{\mathrm{B}}=\eta^{\mathrm{A}}$. 
The shear load is applied through a stress rate $\dot{\tau}$, here given by $\dot{\tau}=2.59\cdot 10^{-5}\,{\widehat{T}^{\mathrm{A}^2}}/B$. This rather high stress rate guarantees numerical stability. \par
\begin{figure}[htbp]
	\centering
	\includegraphics[width=1.0\linewidth]{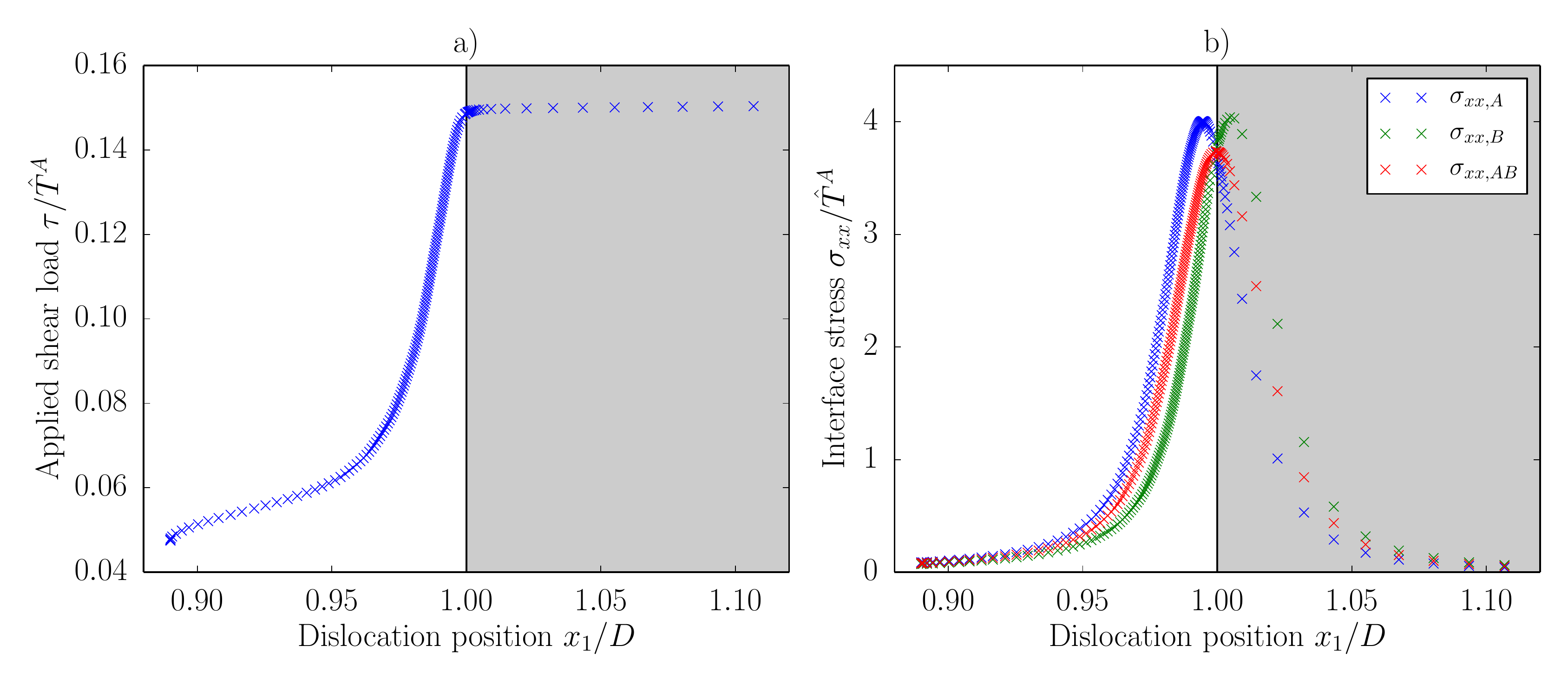}
	\caption{(a) Dislocation position $x_1$ with respect to the applied shear load $\tau$; (b) nodal stresses extrapolated from the element in Phase A/B $\sigma_{xx,A/B}(L_{\mathrm{A}},0)$ and the element averaged nodal stress $\sigma_{xx,AB}$ as a function of the dislocation position.}
	\label{fig:Dislocation_transmission}
\end{figure}

Figure \ref{fig:Dislocation_transmission}a shows the dislocation evolution for $h=b/2$ in terms of the dislocation position $x_1$ (horizontal axis) as a function of the applied shear load $\tau$ (vertical axis). Phase B is visualised by the grey domain. 
After nucleation at $\tau=\tau_{\mathrm{nuc}}$ the dislocation moves under the influence of the applied shear load towards the phase boundary, where it is obstructed. The physics governing the dislocation obstruction stems from the introduced phase contrast itself. The stiffer Phase B induces an image stress field that repels the approaching dislocation. Naturally, the shear load needs to be increased further to overcome the repulsion and to push the dislocation closer to the interface. Eventually, the applied shear load reaches a critical value, the external transmission stress $\tau_{\mathrm{trans}}$, here $\tau_{\mathrm{trans}}\approx 0.15 \,\widehat{T}^{\mathrm{A}}$. At this point, at which the dislocation is situated at the phase boundary ($x_1=L_{\mathrm{A}}$), the local shear stress acting on the dislocation is in equilibrium with the transmission resistance. A small increase of the shear load drives the dislocation rapidly into Phase B. The slow and linear motion of the dislocation prior to transmission resembles quasi-static solutions. It is therefore expected that the transmission process is only slightly influenced by the magnitude of the applied stress rate. \par
To study the local stress build-up during dislocation transmission, the maximum normal stress $\max(\sigma_{xx})$ at the boundary $x=L_{\mathrm{A}}$ is evaluated. Following the analytical solution the maximum of $\sigma_{xx}$ occurs in $\Omega_-$ at $y=0$ \cite{Hirth1982}. This point is shared by two elements, one lying in Phase A and one in Phase B. Hence, two stress evaluations are available at the respective nodal position, extrapolated from the element in Phase A and Phase B, here denoted as $\sigma_{xx,\mathrm{A}}$ and $\sigma_{xx,\mathrm{B}}$. Figure \ref{fig:Dislocation_transmission}b displays these stresses with the elemental averaged nodal stress $\sigma_{xx,\mathrm{AB}} = (\sigma_{xx,\mathrm{A}} + \sigma_{xx,\mathrm{B}})/2$ (vertical axis) as a function of the dislocation position (horizontal axis). Again, the grey area indicates Phase B. The deviation visible in $\sigma_{xx,\mathrm{AB}}$ originates from the spatial discretisation with a coarse element size of $h=b/2$. Clearly, a high interface stress emerges only when a dislocation reaches the direct vicinity of the interface, with its maximum at $x_1=L_{\mathrm{A}}$.\par
\begin{figure}[htbp]
	\centering
	\includegraphics[width=1.0\linewidth]{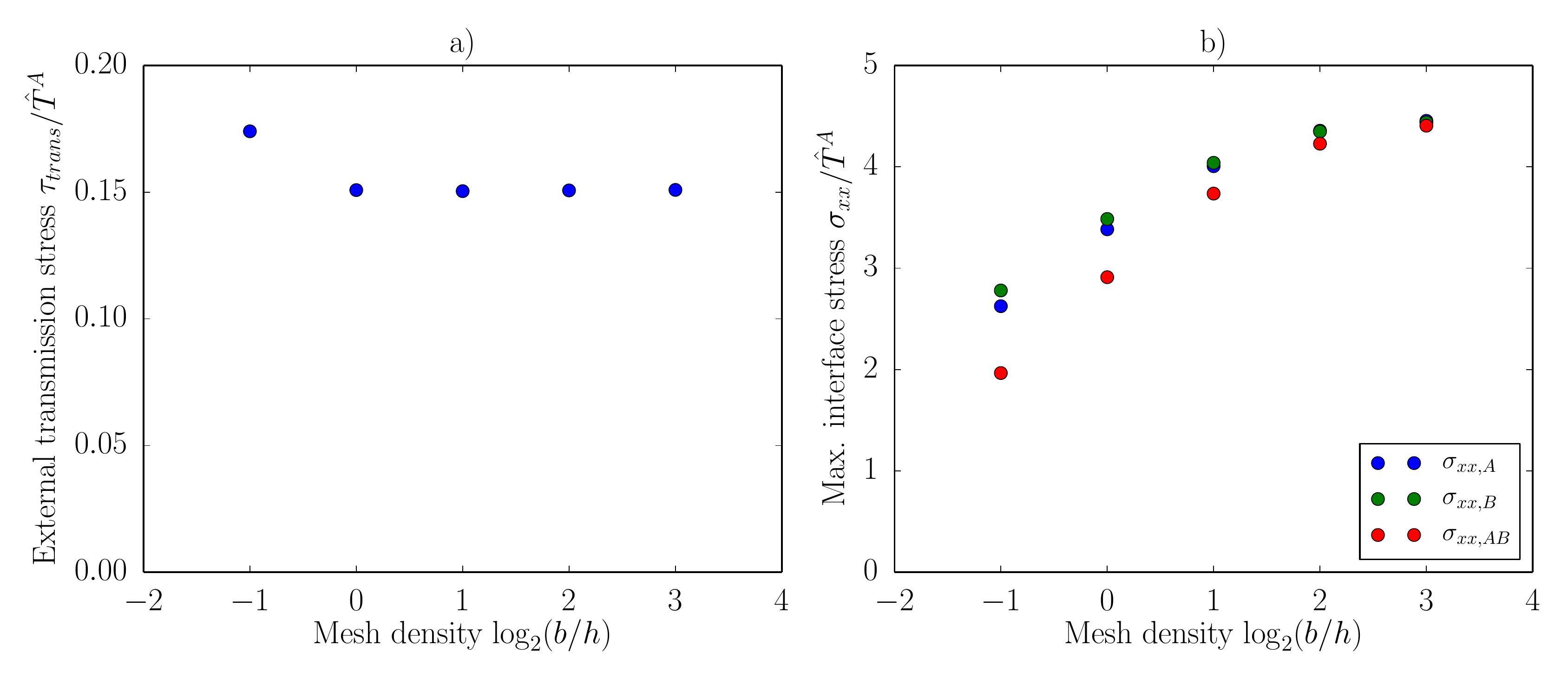}
	\caption{Influence of the mesh size $h$ on (a) the external transmission stress $\tau_{trans}$ and (b) the maximum nodal stresses extrapolated from the element in Phase A/B $\sigma_{xx,A/B}(L_{\mathrm{A}},0)$ and the element averaged nodal stress $\sigma_{xx,AB}$.}
	\label{fig:Mesh-dependency-tau_trans-sxx}
\end{figure}
The mesh convergence results for dislocation transmission are presented in Figure \ref{fig:Mesh-dependency-tau_trans-sxx}. Figure \ref{fig:Mesh-dependency-tau_trans-sxx}a plots the external transmission stress $\tau_{\mathrm{trans}}$ (vertical axis) against the element size $h$ (horizontal axis). Although $h=b$ exhibits an inaccuracy in terms of the equilibrium position (see Figure \ref{fig:Mesh-dependency-disl-pos}) it provides with $\tau_{\mathrm{trans}}\approx 0.15 \,\widehat{T}^{\mathrm{A}}$ a reliable identification of the external transmission stress. In Figure \ref{fig:Mesh-dependency-tau_trans-sxx}b the convergence of the maximum interface stresses $\sigma_{xx,\mathrm{A/B}}(x=L_{\mathrm{A}},y=0)$ and their average $\sigma_{xx,\mathrm{AB}}(x=L_{\mathrm{A}},y=0)$ are depicted as a function of $h$. For a good approximation, an element size of $h=b/8$ is required -- equivalent to the stress field approximation in a homogeneous medium (see Figures \ref{fig:Shear-stress-line_plot} and \ref{fig:Normal-stress-line_plot}). With $\sigma_{xx,\mathrm{A/B}}=4.4\,\widehat{T}^{\mathrm{A}}$ the normal stress clearly exceeds the conventional approximation of the theoretical strength $E/10$. This is due to the subatomic discretisation in the continuum model where the conventional approximation no longer applies.\par

\section{Influence of phase contrast and pile-ups on dislocation transmission}
\label{sect:Results}
\subsection{Transmission in a single dislocation case}
\label{sect:results-single}
As illustrated in Section \ref{sect:Transmission-Verification}, a phase contrast provides a natural source for dislocation obstruction. The aim of this section is to investigate the influence of different phase contrasts on the external transmission stress $\tau_{\mathrm{trans}}$ in a single dislocation framework. For this purpose model dimensions of $L\times 2H = 400b \times 500b$ and $L_{\mathrm{A}} = 200b$ are adopted. The element size is set to $h=b$ which provides adequate results for $\tau_{\mathrm{trans}}$ (see Section \ref{sect:Transmission-Verification}). Material properties are: $\mu^{\mathrm{A}} = 80.77\,\mathrm{GPa}$, $\nu^{\mathrm{B}}=\nu^{\mathrm{A}}=0.3$, $\widehat{T}^{\mathrm{A}}=\mu^{\mathrm{A}}/2\pi$, $b^{\mathrm{B}}=b^{\mathrm{A}}=b=0.25\,\mathrm{nm}$, $\eta^{\mathrm{A}}=\eta^{\mathrm{B}}=B/b$, $B = 10^{-4}\,\mathrm{Pa\,s}$ and $\tau_{\mathrm{nuc}}=0.012\,\widehat{T}^{\mathrm{A}}$. The shear modulus $\mu^{\mathrm{B}}$ and the glide plane traction amplitude $\widehat{T}^{\mathrm{B}}$ are varied in the case studies examined. The shear load is applied at the rate $\dot{\tau}=0.32\cdot 10^{-5}\,{\widehat{T}^{\mathrm{A}^2}}/B$.\par
\begin{figure}[htbp]
	\centering
	\includegraphics[width=0.5\linewidth]{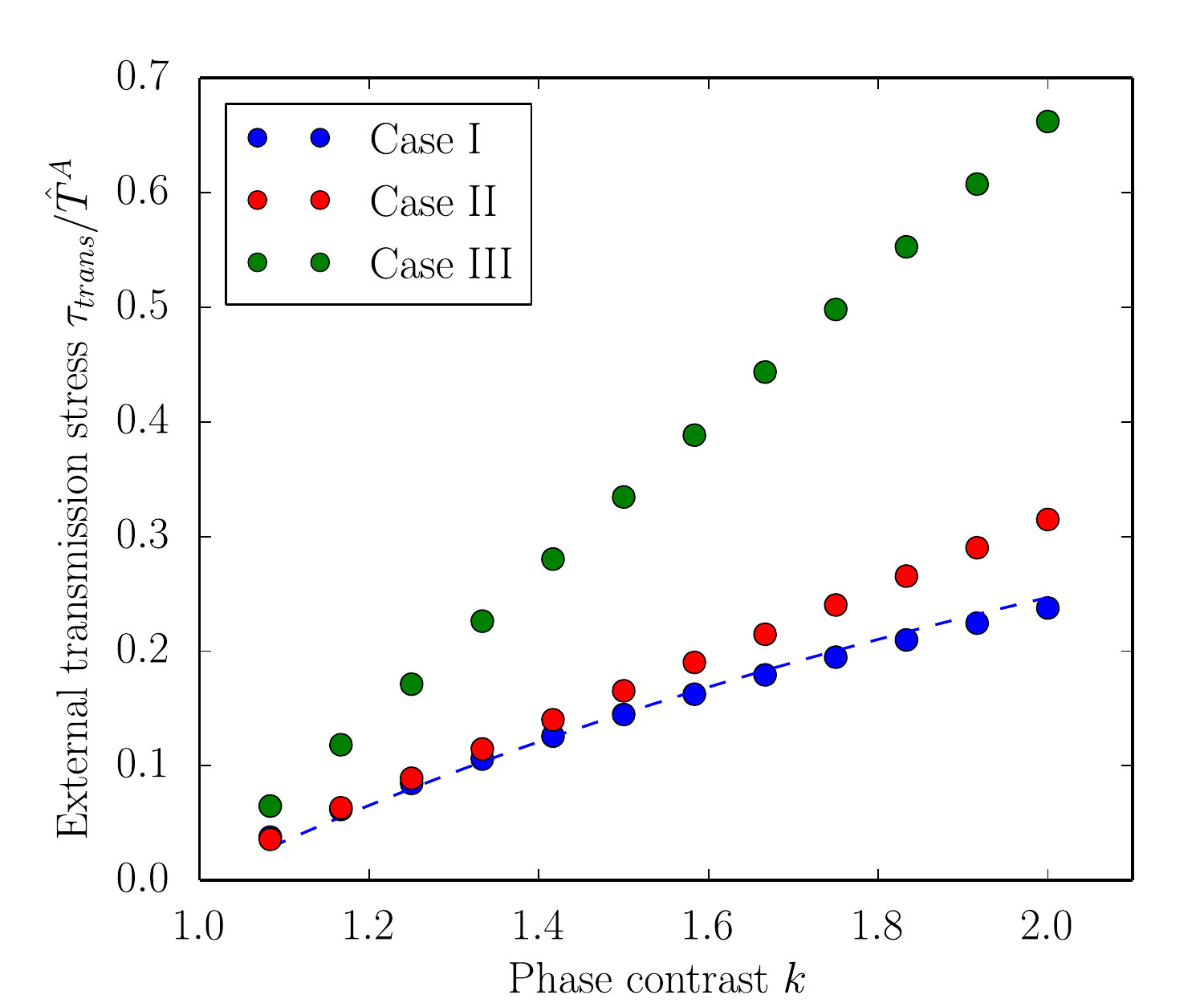}
	\caption{Influence of phase contrast $k$ on the external transmission stress for Case I: $\mu^B=k\mu^A$ and $\widehat{T}^B=\widehat{T}^A$; Case II: $\mu^B=\mu^A$ and $\widehat{T}^B=k\,\widehat{T}^A$; Case III: $\mu^B=k\mu^A$ and $\widehat{T}^B=k\,\widehat{T}^A$. The dashed line represents the scaling factor for the image stress of an edge dislocation on itself $\tau_{xy}\cdot(x-L_{\mathrm{A}})/b$ \cite{Head1953-Edge}}
	\label{fig:phase-contrast}
\end{figure}

In terms of phase contrast, three selected cases are considered ($k>1$):
\begin{enumerate}[leftmargin=1.75cm, labelsep=*, align=left, itemsep=-0.1cm, label = {Case} \Roman*:, topsep=0pt]
	\item $\mu^{\mathrm{B}} = k\mu^{\mathrm{A}}$ and $\widehat{T}^{\mathrm{B}}=\phantom{k}\,\widehat{T}^{\mathrm{A}}$
	\item $\mu^{\mathrm{B}}=\phantom{k}\mu^{\mathrm{A}}$ and $\widehat{T}^{\mathrm{B}}=k\,\widehat{T}^{\mathrm{A}}$
	\item $\mu^{\mathrm{B}}=k\mu^{\mathrm{A}}$ and $\widehat{T}^{\mathrm{B}}=k\,\widehat{T}^{\mathrm{A}}$
\end{enumerate}
These cases are chosen such that two different sources of dislocation obstruction are triggered. One relates to the repelling long-range image stress field induced by the difference in elastic properties. The second source stems from an increased traction amplitude $\widehat{T}$ in Phase B which causes a short-range image stress field. The results are shown in Figure \ref{fig:phase-contrast} in terms of the external transmission stress $\tau_{\mathrm{trans}}$ as a function of the phase contrast $k$. \par
For the pure elasticity contrast, in Case I, a trend is obtained that is comparable to the scaling factor for the image stress of an edge dislocation at $x$ on itself as derived for a Volterra dislocation by Head \cite[Section 10]{Head1953-Edge}. This trend is included in Figure \ref{fig:phase-contrast} as a dashed curve. Despite its premise of a Volterra dislocation it presents a fairly good approximation for $\tau_{\mathrm{trans}}^{i}$. 
Both Cases II and III exhibit a linear trend between $\tau_{\mathrm{trans}}$ and $k$ with $\tau_{\mathrm{trans}}^{\mathrm{III}}>\tau_{\mathrm{trans}}^{\mathrm{II}}>\tau_{\mathrm{trans}}^{i}$. This shows that for a single dislocation the glide plane contrast and the related repulsive short-range image stress field has a higher impact on dislocation obstruction than the pure elasticity contrast with its long-range image stress field. Furthermore, since $\tau_{\mathrm{trans}}^{\mathrm{III}}>\tau_{\mathrm{trans}}^{\mathrm{II}}+\tau_{\mathrm{trans}}^{i}$, the influence of glide plane and elasticity contrast cannot be decoupled, i.e., the elasticity contrast increases the short-range stress contribution that is invoked by the higher traction amplitude.\par
%
%
\subsection{Transmission in a dislocation pile-up case}
So far the case of a single dislocation dipole was considered. This is next extended to multiple dislocations, allowing for the formation of dislocation pile-ups, which will impact dislocation transmission. Model settings are used as in the single dislocation case. Phase contrasts are adopted in alignment with Cases I-III, as defined in Section \ref{sect:results-single}, and $k=2$.\par
\begin{figure}[htbp]
	\centering
	\includegraphics[width=0.7\linewidth]{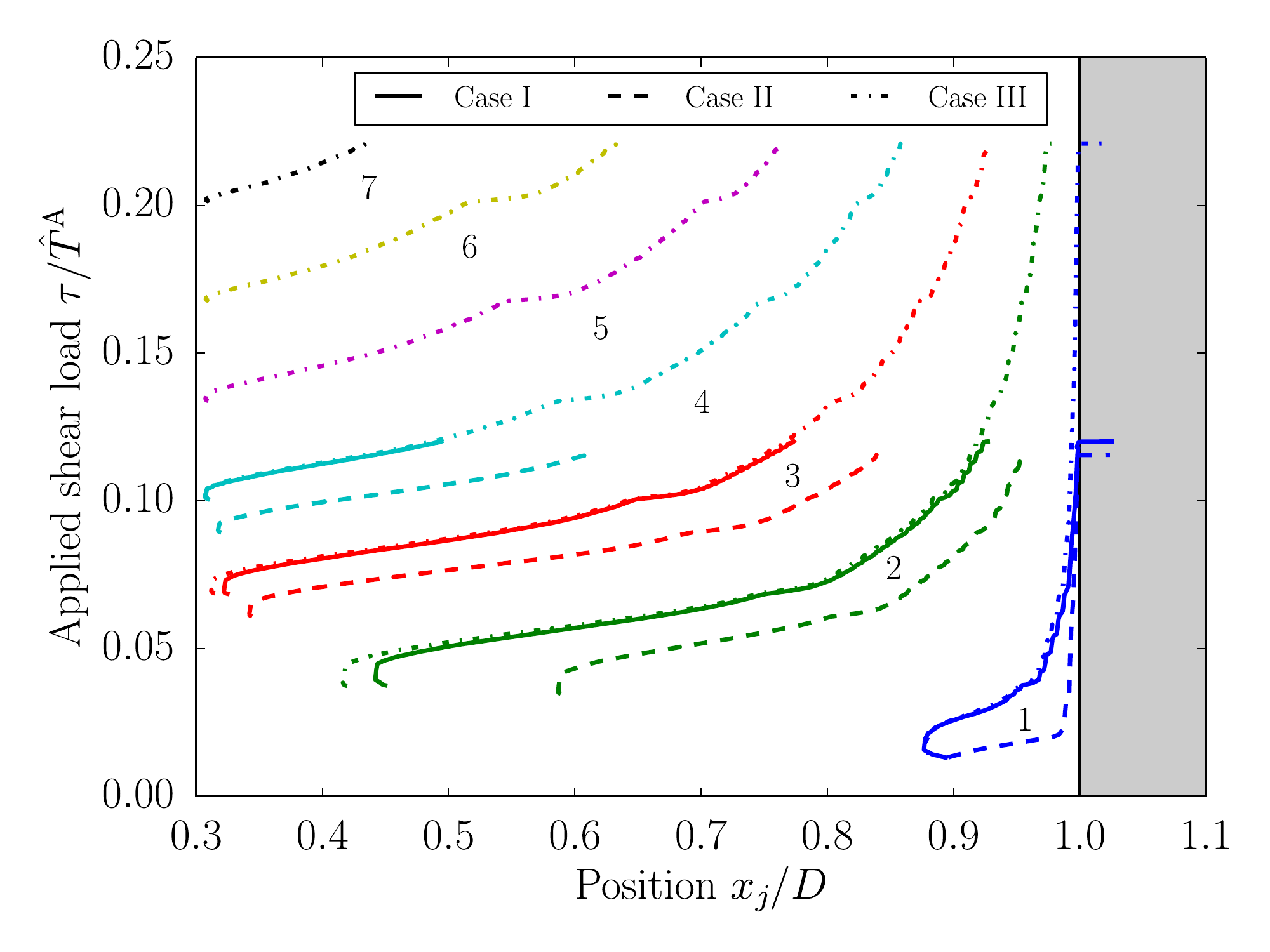}
	\caption{Dislocation position with respect to the applied shear load for each dislocation for Case I: $\mu^B=2\mu^A$ and $\widehat{T}^B=\widehat{T}^A$ (solid line); Case II: $\mu^B=\mu^A$ and $\widehat{T}^B=2\,\widehat{T}^A$ (dashed line); Case III: $\mu^B=2\mu^A$ and $\widehat{T}^B=2\,\widehat{T}^A$ (dash-dot line).}
	\label{fig:pile-up-evolution}
\end{figure}

 
During the evolution process dislocations are nucleated at positions following from Section \ref{sect:disl_nuc}. Figure \ref{fig:pile-up-evolution} displays the dislocation evolution and the eventually triggered dislocation transmission for the different phase contrasts. No upper limits for the number of dislocations are used. The position of each individual dislocation $x_{j}$ (horizontal axis), where $\Delta = (j-1/2)b$, is plotted by a single line against the applied shear load $\tau$ (vertical axis). \par
Upon increasing the applied shear load, dislocations are nucleated one by one and pile up at the phase boundary. Here, the absence of a long-range image stress field in Case II becomes apparent as the leading dislocation moves directly into the phase boundary. By further increasing the shear load, the leading dislocation gets ultimately transmitted into Phase B. Transmission is recorded for Case I with 4 dislocations at $\tau_{\mathrm{trans}}\approx 0.120\, \widehat{T}^{\mathrm{A}}$, for Case II with 4 dislocations at $\tau_{\mathrm{trans}} \approx 0.116\, \widehat{T}^{\mathrm{A}}$ and for Case III with 7 dislocations at $\tau_{\mathrm{trans}} \approx 0.221\, \widehat{T}^{\mathrm{A}}$. The nucleation of dislocations in the vicinity of existing dislocations triggers a momentarily strong driving force on the current pile-up -- i.e., a small kink in the remaining dislocation evolution curves. \par
In all cases, the phase contrast and the grain size $2L_{\mathrm{A}}$ induce an upper limit on the number of dislocations in the pile-up. While the phase contrast defines the resistance against dislocation transmission and relates hence to the external transmission stress $\tau_{\mathrm{trans}}$, the grain size limits the number of stable dislocation dipoles that can be accommodated in Phase A at $\tau=\tau_{\mathrm{trans}}$. For Case II, where no elasticity related long-range image stress field is present, an analytical expression for this upper limit can be found. Following \cite{Hirth1982} the number $N$ of positive dislocations in a dipole pile-up of length $l$ within a homogeneous medium reads:
\begin{equation}
\label{eq:dipole-length}
N=\frac{(1-\nu)l\tau}{\mu b}
\end{equation}
with a resolved shear stress acting on the leading dislocation of
\begin{equation}
\label{eq:leading-stress}
\tau^{\mathrm{res}}_1 = \frac{\pi}{4}N\tau
\end{equation}
Combining Eq. \eqref{eq:dipole-length} and \eqref{eq:leading-stress} yields
\begin{equation}
\label{eq:number-disl}
N = \sqrt{\frac{4}{\pi}\frac{(1-\nu)l\tau^{\mathrm{res}}_1}{\mu b}}
\end{equation}
Projecting this relation upon the presented model, the natural upper limit of $N$ can be approximated when a dipole pile-up is fully formed at the point of transmission. Accordingly, $\tau_1^{\mathrm{res}}$ is taken equal to $\tau_{\mathrm{trans}}$ of a single dislocation and $l$ is the grain size of Phase A, $2L_{\mathrm{A}}$. In the present case $N=4$ results (after rounding down), which is in agreement with the simulation. If the size of Phase A is so large that transmission occurs before the dipole pile-up is fully formed, a mixture between single pile-up and dipole pile-up is captured. A straightforward analytical expression is then not possible. An analytical expression also fails for Cases I and III where the presence of dislocations gives rise to the repulsive long-range image stress field and hence an increased dislocation obstruction.\par
\begin{figure}[htbp]
	\centering
	\includegraphics[width=0.5\linewidth]{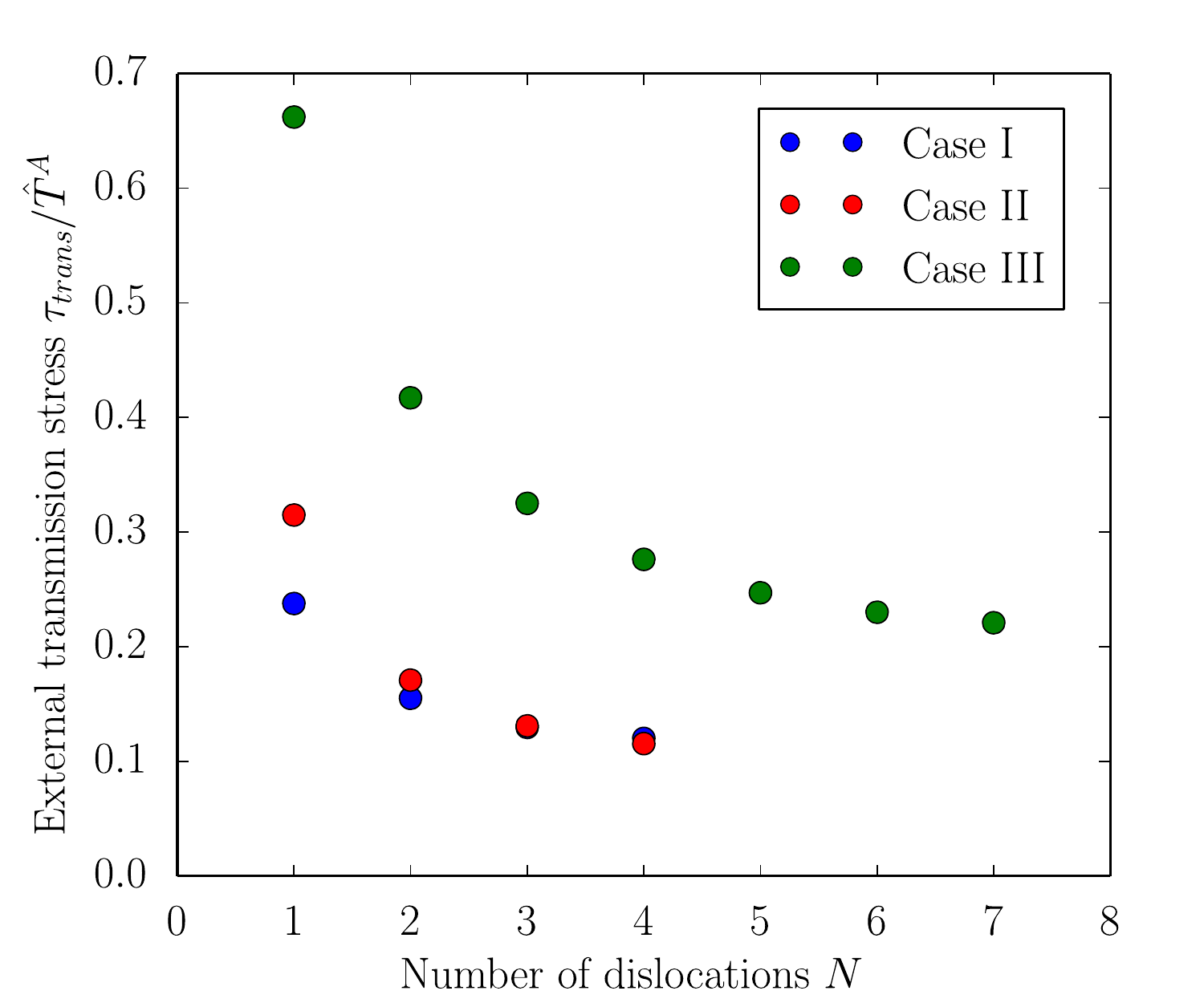}
	\caption{External transmission stress with respect to the number of dislocations for Case I: $\mu^B=2\mu^A$ and $\widehat{T}^B=\widehat{T}^A$; Case II: $\mu^B=\mu^A$ and $\widehat{T}^B=2\widehat{T}^A$; Case III: $\mu^B=2\mu^A$ and $\widehat{T}^B=2\widehat{T}^A$.}
	\label{fig:tau-trans_vs_N-disl}
\end{figure}

The influence of the number of dislocations on the external transmission stress $\tau_{\mathrm{trans}}$ is studied by varying the number of dislocations $N\le 4$ for Cases I and II and $N\le 7$ for Case III. The results are presented in Figure \ref{fig:tau-trans_vs_N-disl}. Although a strong decay in external transmission stress is obvious, the expected reduction towards zero is not apparent. The reason is twofold. i) The Dirichlet boundary conditions \eqref{eq:boundary_value1}-\eqref{eq:boundary_value3} are applied under the assumption of linear elasticity, neglecting the influence of slip along $\Gamma_{\mathrm{gp}}$. While this effect is marginal for a single dislocation, its impact grows with the number of dislocations. Thus, $\tau_{\mathrm{trans}}$ is increasingly overestimated. ii) At the point of transmission, the dislocation pile-ups are not in their static equilibrium, yet. The resolved shear stress on the leading dislocation is lower than for a fully developed dislocation pile-up. Hence, for dislocation transmission the applied shear load needs to be increased further than expected for a fully developed pile-up.\par
Comparing Cases I and II, a growing influence of the elasticity induced long-range image stress field becomes obvious. As more dislocations are introduced, the repulsive long-range image stress field stacks up. Consequently, dislocation obstruction gets stronger than for the pure glide plane contrast where dislocation obstruction arises from the short-range image stress field caused by solely the transmitting dislocation. Hence, Case I exhibits a smaller decay in external transmission stress than Case II. 
%
%
\section{Conclusion}
\label{sect:Conclusion}
A numerical 2D model for edge dislocation--phase boundary interaction has been presented. It consists of two linear-elastic phases with an embedded glide plane according to the Peierls--Nabarro model. To solve the governing equations, the model is discretised spatially by FEM and in time by a backward Euler scheme. This approach provides a natural interplay between dislocations, phase boundary and external boundary conditions with only a limited number of parameters. It is therefore a suitable method for studying the stress build-up at phase boundaries due to dislocation pile-ups.\par
This chapter considered a soft Phase A that is flanked by a harder Phase B. At the centre of Phase A, a dislocation source emits edge dislocation dipoles under a sufficiently high local shear stress. A single glide plane, perpendicular to a fully coherent and non-damaging phase boundary, was analysed. A shear load was applied in the form of Dirichlet boundary conditions. \par 
To verify the presented model, the stress field of a single dislocation within a homogeneous material was studied. It could be shown that for a sufficiently fine discretisation ($h\le b/8$), the numerical solution converges towards the analytical solution. Similarly, the adequate spatial discretisation for capturing dislocation motion and dislocation transmission in a two-phase microstructure was determined, with $h\le b/2$.\par 
For a first qualitative study, the influence of phase contrast and the number of dislocations in a dipole pile-up on dislocation transmission was determined. Due to the presence of a phase contrast, a resistance against dislocation motion -- from soft to hard phase -- emerges. Different sources of resistance were identified as i) a long-range image stress field due to the difference in elastic properties and ii) a short-range image stress field arising from the increased traction amplitude of the PN model. \par
In order to differentiate between the obstruction due to the long- and short-range stress fields, a transmission study was performed exploring three different kinds of phase contrast: pure phase contrast in elastic properties, pure phase contrast in the glide plane traction amplitude and a combination of both.
The following insights were gained. i) Dislocation obstruction due to pure elasticity contrast scales in a comparable manner as the image shear stress of a Volterra edge dislocation on itself as derived by Head \cite{Head1953-Edge}. ii) For a pure contrast in the PN traction amplitude, obstruction due to the short-range image stress field scales linearly and is stronger than obstruction due to a pure long-range image stress field. iii) For combined phase contrasts the short-range image stress field is enhanced by the elasticity contrast.\par
A study of the influence of the number of dislocations in a dipole pile-up followed. An increasing number of dislocations strongly decreases -- as expected -- the applied shear load needed for dislocation transmission. A pure elasticity contrast exhibited a smaller decrease in external transmission stress than a pure glide plane contrast. This confirms an enhanced repulsive long-range stress field for an increasing number of dislocations. For all phase contrasts the external transmission stress did not converge asymptotically towards zero as expected from the theory. Two reasons could be identified as follows. i)
The Dirichlet boundary conditions neglected the influence of slip along the glide plane. Accordingly, the obtained external transmission stress was somewhat overestimated, an effect that increases with a the number of dislocations. A larger model domain would reduce this effect. ii) At the point of transmission, the dislocation pile-ups were not in their static equilibrium yet. For fully evolved dislocation pile-ups, the external transmission stress is expected to drop -- an effect that increases with the number of dislocations.\par
Despite the simplifications made, the model is a suitable tool to gain insight into edge dislocation--phase boundary interaction in two-phase microstructures. Its strength is in particular that it incorporates the repulsion of dislocations by the phase boundary and the resistance against dislocation transmission exclusively through the adopted mechanics. No additional constitutive relations for dislocation--interface interactions, e.g., a transmission criterion, are required -- cf. DDD where this would be necessary. On the other hand, the model can be better controlled and is computationally significantly less expensive than MD.
%
%
\section*{Acknowledgements}
We would like to thank Franz Roters and Pratheek Shanthraj of the Max Planck Institute for Iron Research for useful discussions on the present work. This research is supported by Tata Steel Europe through the Materials innovation institute (M2i) and Netherlands Organisation for Scientific Research (NWO), under the grant number STW 13358 and M2i project number S22.2.1349a.
%
%

%
%

\begin{appendices}
\section{Determination of the damping coefficient}
\label{Appendix:Damping}
A suitable value for the damping coefficient $\eta$ in Eq. \eqref{eq:PN-num} can be found through a comparison with DDD. Generally, a dislocation that travels the distance of a Burgers vector $b$ dissipates the work
\begin{equation}
\label{eq:dis-general}
W_{\mathrm{dis}}=\int_{x_{j}}^{x_{j}+b} F(x)\,\mathrm{d}x
\end{equation}
where $x_{j}$ and $F(x)$ denote the initial dislocation position and the force acting on the dislocation, respectively. For DDD the force $F(x)$ equates the Peach--Koehler force $F(x) = \tau^{\mathrm{res}}b$. The resolved shear stress $\tau^{\mathrm{res}}$ acting on the dislocation is generally related to the dislocation velocity $v$ via a linear drag law
\begin{equation}
\tau^{\mathrm{res}} = \frac{B}{b}v
\end{equation}
with the drag coefficient $B$. Assuming a constant $\tau^{\mathrm{res}}$ over the distance $b$ the dissipated energy equals
\begin{equation}
W_{\mathrm{dis}}^{\mathrm{DDD}} = Bvb
\end{equation}
Within the Peierls--Nabarro model there is no point force such as the Peach--Koehler force in DDD anymore. The force $F(x)$ is now calculated by
\begin{equation}
\label{eq:force-PN}
F(x)=\int_{x_{j}^*-\infty}^{x_{j}^*+\infty}\tau^{\mathrm{res}}(x)\,\mathrm{d}x
\end{equation}
where $x_{j}^*$ is the current dislocation position. Adopting the shear traction-disregistry law of Eq. \eqref{eq:PN-num} for $\tau^{\mathrm{res}}(x)$ the force $F(x)$ can be approximated for a single phase model by
\begin{equation}
F(x)=\int_{x_{j}^*-\infty}^{x_{j}^*+\infty}\eta\dot{\Delta}(x)\,\mathrm{d}x
\end{equation}
Modifying the analytical solution \eqref{eq:disreg-analytical} for an arbitrary dislocation position $x_{j}$
\begin{equation}
\Delta(x) = -\frac{b}{\pi}\tan^{-1}\left(\frac{x-x_{j}^*}{\zeta}\right)+\frac{b}{2}
\end{equation}
with 
\begin{equation}
\dot{\Delta}(x)=\frac{\mathrm{d}\Delta}{\mathrm{d}x}v
\end{equation}
one obtains for a constant dislocation velocity $v$ the force
\begin{equation}
F(x) = \eta b v
\end{equation}
and the dissipation
\begin{equation}
W_{\mathrm{dis}}^{\mathrm{PN}}=\eta b^2 v
\end{equation}
Requiring the dissipated energy in both approaches to be equal gives
\begin{equation}
\eta = \frac{B}{b}
\end{equation}
which is an adequate approximation for the damping coefficient.
\end{appendices}
\end{document}